\documentclass[preprint2]{aastex}

\usepackage{graphicx}
\usepackage{psfig}
\slugcomment{Accepted for publication in ApJ Letters}

\shorttitle{The radio-luminosity -- black hole mass correlation}
\shortauthors{Lacy et al.}

\begin{document}


\title{The radio-luminosity -- black hole mass correlation for quasars
from the FIRST Bright Quasar Survey, and a ``unification scheme''
for radio-loud and radio-quiet quasars.}


\author{Mark Lacy\altaffilmark{1,2}, 
Sally A.\ Laurent-Muehleisen\altaffilmark{2}, 
Susan E.\ Ridgway\altaffilmark{3}, 
Robert H.\ Becker\altaffilmark{2} and Richard L.\ White\altaffilmark{4}}
\altaffiltext{1}{IGPP, L-413, Lawrence Livermore National Laboratory, 
Livermore, CA~94550; mlacy@igpp.ucllnl.org}
\altaffiltext{2}{Department of Physics, University 
of California, 1 Shields Avenue, Davis, CA 95616; slauren@igpp.ucllnl.org,
bob@igpp.ucllnl.org}
\altaffiltext{3}{Department of Physics and Astronomy, Johns Hopkins 
University, Baltimore, MD~21218; ridgway@pha.jhu.edu}
\altaffiltext{4}{Space Telescope Science Institute, Baltimore, MD~21218;
rlw@stsci.edu}

\begin{abstract}
Several independent lines of evidence now point to a correlation between 
black hole mass, $M_{\rm bh}$, and radio-luminosity. In this paper we discuss 
the correlation for quasars from the FIRST Bright Quasar Survey
(FBQS), using black 
hole mass estimates from H$\beta$ linewidths. The FBQS objects
fill in the gap between the radio-loud and radio-quiet quasars
in the radio-luminosity -- optical-luminosity plane, and we find that they 
fill the corresponding gap in the $M_{\rm bh}$ -- radio luminosity correlation.
There is thus a continuous variation of radio
luminosity with $M_{\rm bh}$, and no evidence for a ``switch'' at some set of 
critical parameter values which turns on powerful radio jets. 
By combining the FBQS data with that for quasars from the Palomar-Green survey
we find evidence for a dependence of radio-luminosity on accretion rate 
relative to the Eddington limit, $L/L_{\rm Edd}$, as well as on $M_{\rm bh}$,
consistent with the well-known radio-optical correlation for radio-loud 
quasars. We therefore suggest a new scheme to ``unify'' radio-loud and 
radio-quiet objects in which radio luminosity scales 
$\propto M_{\rm bh}^{1.9\pm 0.2}(L/L_{\rm Edd})^{1.0}$
for $L/L_{\rm Edd}\sim 0.1$, with an apparently weaker accretion rate 
dependence at low $L/L_{\rm Edd}$. The scatter about this relation is 
$\pm 1.1$ dex, and may well hide significant contributions from other 
physical effects, such as black hole spin and radio source environment.

\end{abstract}



\section{Introduction}

Evidence for a link between black hole mass and radio-loudness in quasars
has been accumulating for the past decade (see Laor 2000), but recently 
several new pieces of evidence have come to light. The 
host galaxy properties suggest a link; nearly all radio-loud quasars and 
radio galaxies are hosted by giant ellipticals, whereas radio-quiet quasars 
can exist in spirals (Bahcall et al.\ 1997; McLure et al.\ 1999). Coupled
with the correlation of black hole mass and bulge mass seen in nearby galaxies
(e.g.\ Magorrian et al.\ 1997), this implies that the radio-loud quasars
and radio galaxies are likely to be hosted by objects with the most massive 
black holes (McLure et al.\ 1999). Although controversial, it seems that 
the fraction of radio-loud quasars does increase with optical luminosity
[Hooper et al.\ 1995; Goldschmidt et al.\ (1999); Impey \& Petry (2000); but 
see Stern (2000) for a dissenting view], again consistent with the idea that  
the most massive black holes are more likely to produce powerful radio 
sources (Lacy, Ridgway \& Trentham 2000). The energy stored in  
giant radio lobes is $\stackrel{>}{_{\sim}} 5\times 10^5 M_{\odot}c^2$; 
assuming a plausible efficiency of conversion, this implies a black hole mass 
$\stackrel{>}{_{\sim}} 10^7M_{\odot}$ (Rawlings \& Saunders 1991). 
Most recently, Laor (2000) estimated black hole masses for $z<0.5$ quasars in 
the optically-selected Palomar-Green (PG) Survey of Green, Schmidt \& Liebert 
(1986), using black hole masses derived from the linewidths of broad H$\beta$ 
and an estimate of the size of the broad-line region. He found a 
significant difference in the 
distributions of black hole masses for radio-loud and radio-quiet quasars, 
with radio-loud quasars generally having more massive black holes.
A correlation between radio luminosity and black hole masses also estimated 
from linewidths, but using the narrow [O{\sc iii}]5007 line, was found by 
Nelson (2000). His correlation 
only worked for radio-quiet quasars, however, perhaps because radio-loud 
quasars tend to have a larger amount of very extended ($\sim 30$ kpc
scale) [O{\sc iii}] emission (Stockton \& MacKenty 1987).

In this paper we discuss the black hole mass -- radio-luminosity correlation
with particular reference to quasars in the FIRST Bright Quasar Survey (FBQS;
Gregg et al.\ 1996; White et al.\ 2000). The unique radio-optical 
selection criteria of this 
survey make it particularly sensitive to quasars with radio luminosities
between the traditional radio-loud and radio-quiet classes. It is thus ideal
for a quantitative investigation of the correlation of black hole mass and 
radio-loudness. We follow Laor (2000) in using estimates of
black hole masses based on H$\beta$ linewidths. McLure \& Dunlop (2000)
show that black holes masses derived in this way
follow the same correlation of black hole mass with host bulge 
luminosity as those derived from stellar dynamics, so are probably 
reasonably accurate.

We assume a cosmology with $H_0=50{\rm kms^{-1}Mpc^{-1}}$ and 
$\Omega=1$, $\Lambda=0$ throughout.

\section{The FBQS sample}

The FBQS
consists of optically unresolved, bright ($R\le$17.8\,mag), blue
($B-R\le$2.0) quasars selected from the FIRST radio survey (Becker, White \&
Helfand 1995).  At the time of the radio observations used herein, the survey
consisted of 607 quasars, 80\% of which were previously
unknown.  The sample used here is over 90\% complete and differs from the
full FBQS sample in two ways: we limited the range of redshifts to be $0.3 <
z < 0.5$ and also only used sources with $M_B < -23$. The upper redshift
limit ensures that H$\beta$ is present in the spectrum. The absolute 
magnitude limit ensures that contamination of the image by host galaxy
light is small [the host galaxies observed by Bahcall et al.\ (1997) 
all have $M_{B}\stackrel{>}{_{\sim}} -22.5$]. $M_B \approx -23$ 
corresponds to our optical selection limit at $z\sim 0.3$, so we have 
excluded lower redshift objects from the sample to minimize the number of 
objects where host galaxy light can contribute significantly to $M_B$. 
The absolute magnitude limit also ensures that our sample should be 
free of any significant biases due to 
exclusion of objects through the FBQS selection criteria, which
requires the image to be stellar on at least one of the Palomar Observatory 
Sky Survey plates. This selection results 
in a sample of 60 FBQS objects.
%, consisting of 26 flat-spectrum quasars 
%and 34 steep-spectrum quasars. 

The sources were observed at 20 and 3.6\,cm (to obtain simultaneous radio
spectral indices, although see below) on 1999 April 12, 16, 18 and 23, with
the VLA in D-array.  Data were reduced in the standard manner
using the AIPS package.  
%Fluxes were measured using the routine
%JMFIT and radio spectral indices calculated assuming $\alpha_{\rm r} = -1.293
%\log(S_{20}/S_{3.6}$ where $S_{20}$ and $S_{3.6}$ are the peak 20\,cm and
%3.6\,cm fluxes, respectively.  
The resolution and baseline coverage of the VLA in  
D-array at 3.6\,cm is close to that of the 20\,cm FIRST data; thus 
extended emission might be detectable in the D-array 20cm maps that was 
missing in the 3.6cm maps. For sources with significant extended emission
evident from the FIRST maps or from the D-array fluxes we 
assumed the extended emission had 
a radio spectral index $\alpha_{\rm r} = 0.75$ (defined such that 
flux density $S_{\nu} \propto \nu^{-\alpha_{\rm r}}$). We then calculated the
integrated spectral index using the total (D-array) flux at 20cm and the 
sum of the peak and estimated extended flux  at 3.6cm. 

Fig.\ 1 shows the optical-luminosity -- radio-luminosity plane for the FBQS
quasars, together with $z<0.5$ quasars from the PG sample 
[whose H$\beta$ FWHM were measured by Boroson \& Green (1992)]. The PG 
quasars are selected at a brighter optical flux limit than the FBQS quasars 
($B\stackrel{<}{_{\sim}}16$), but as 
many of them are at $z<0.3$ there is a reasonable overlap in optical luminosity
between the two samples. As expected, the FBQS objects fill in the region 
between the radio-loud PG quasars (with radio luminosities at 5GHz, 
$L_{\rm 5GHz}\stackrel{>}{_{\sim}} 10^{24.5} {\rm WHz^{-1}sr^{-1}}$)
and the radio-quiet PG quasars 
(with $L_{\rm 5GHz}\stackrel{<}{_{\sim}}10^{23} {\rm WHz^{-1}sr^{-1}}$) 
\footnote{These divisions between radio-loud, 
radio-quiet and radio intermediate
quasars are used throughout this paper.}.

The samples have been divided into steep spectrum ($\alpha_{\rm r} \geq 0.5$) 
and flat spectrum ($\alpha_{\rm r}<0.5$) subsamples
(for some radio-faint PG quasars no radio spectral index information was 
available; they have been assumed to have $\alpha_{\rm r}=0.6$). 

To measure the FWHM of H$\beta$ we first subtracted the continuum and, where
necessary, an Fe{\sc ii} profile (Boroson \& Green 1992) and 
[O{\sc iii}]4959+5007 emission. As the quality of the spectra varied widely, 
FWHM were estimated by averaging those obtained from fits of Gaussian and 
Lorentzian profiles. This procedure was adopted as, where the signal-to-noise
was high enough to make direct measurements of the FWHM, they typically 
lay between the values from Gaussian and Lorentzian fits. The error in a 
typical measurement is $\approx 10\%$, which translates to an $\approx 20\%$
error in black hole mass. We made a small correction to allow for 
the orientation dependence of FWHM H$\beta$ (e.g.\ Brotherton 1996) by a 
factor of $R_{\rm c}^{0.1}$, where $R_{\rm c}$ is the ratio of core to 
extended radio fluxes (where $R_{\rm c}$ could not be measured we assumed 
$R_{\rm c}=0.1$ for steep-spectrum sources and $R_{\rm c}=10$ for flat-spectrum
ones).

\section{Results}

\begin{figure}[ht]
\plotone{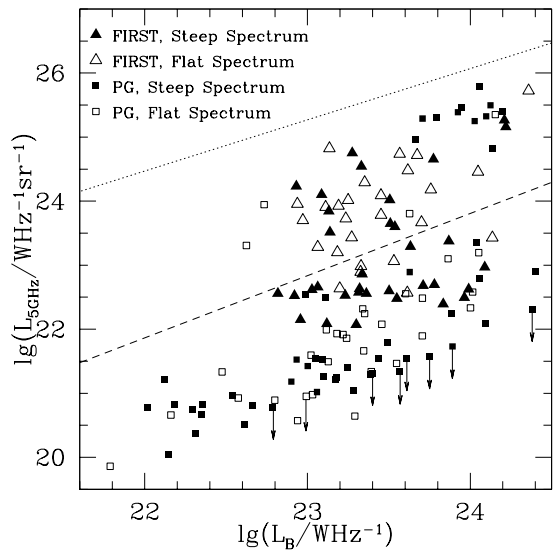}
\caption{The radio-luminosity -- optical luminosity plane for the samples
of quasars discussed in the text. The dotted line is the radio-optical
correlation for radio-bright quasars from the MRC sample
(Serjeant et al.\ 1998). The dashed line corresponds to a radio-loudness
parameter of $R^*=10$, the traditional division between radio-loud and 
radio-quiet objects.}
\end{figure}

\begin{figure}[ht]
\plotone{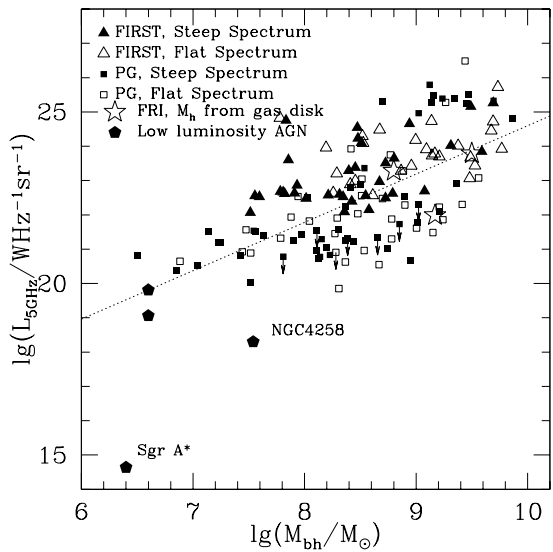}
\caption{Radio luminosity versus black hole mass. The dotted line is the 
best fit to the correlation measured using the PG and FBQS quasars.}
\end{figure}

We derive black hole masses ($M_{\rm bh}$)
following the prescription of Laor (1998), but use the recent 
empirical determination 
of the broad-line region radius from Kaspi et al.\ (2000). We
plot them against radio luminosity in Fig.\ 2. The correlation of black 
hole mass with radio luminosity suggested by the work of Franceschini, 
Vercellone \& Fabian
(1998) and Laor (2000) is clearly present in the data. For the FBQS objects
{\em alone} the Spearman rank correlation coefficient is 0.52 for the 60
objects in the sample, with a probability that no correlation is present of
0.01\%, taking the 34 steep spectrum objects only this is 0.48 
with a probability of no correlation of 0.6\%. When the PG quasars
are included the correlation becomes stronger still.

The FBQS objects fill
in the gap between the radio-loud and radio-quiet objects in Fig.\ 1 of 
Laor (2000), providing evidence of a continuous dependence of radio luminosity
on black hole mass. This suggests that selection effects, rather than 
underlying physics, could be responsible for producing the radio-quiet -- 
radio-loud quasar dichotomy seen in some optically-selected samples (e.g.\
the PG sample). We speculate 
that the selection methods used for the FBQS have highlighted the 
radio-intermediate population, which for some reason is not so common 
in samples selected purely on the basis of radiative luminosity from the 
accretion disk. These issues 
will be discussed further in a future paper (Brotherton et al.\ 2001). We
merely comment here that the FBQS is complete for radio-loud 
and radio-intermediate quasars, thus
the ratio of radio-loud to radio-intermediate quasars in the FBQS should
be representative of that of the quasar population as a whole.

We also plot some galaxies with independent black hole
mass estimates: three FRI galaxies 
with detected optical nuclei, M~84, M~87 and NGC~6251,  
whose black hole masses have been measured through the kinematics of their 
nuclear gas disks (e.g.\ Ferrarase \& Ford 1999), three objects 
with low-mass black holes from the sample of Ho (1996) (including NGC~4258, 
whose black hole mass has been accurately measured using maser emission),
and Sgr~A*, the source at the Galactic Center [using data from Falcke \& 
Biermann (1999)]. The slope of the correlation of 
${\rm lg}L_{\rm 5GHz}$ against ${\rm lg}M_{\rm bh}$ measured for the steep
spectrum quasars only using the {\sc emmethod} task in {\sc iraf}
(Isobe, Feigelson \& Nelson 1986), which 
takes into account limits in some of the data, is $1.4\pm 0.2$. As expected, 
the flat-spectrum objects in each sample generally plot above the steep 
spectrum objects, but the mean enhancement due to Doppler boosting is, on
average, not large (only $\approx 60$\% in the FBQS sample). It is possible
that radio emission from star-formation processes could add significantly
to the radio luminosities, particularly for the low radio-luminosity objects, 
which would tend to make our estimate of the slope low. However, Miller, 
Rawlings \& Saunders (1993) argue that most of the radio emission in most PG 
quasars is related to the AGN.

Fig.\ 3 shows the correlation of the radio-loudness parameter, $R^*$, 
the ratio of rest-frame radio and optical luminosities as defined by 
Sramek \& Weedman (1980), with $M_{\rm bh}$. A Spearman correlation analysis
[modified for use with data containing limits (Isobe et al.\ 1986)] using
the steep spectrum objects only shows that the probability of this correlation
arising by chance is $\sim 10^{-4}$. 
%This correlation suggests 
%that radio and optical luminosity have very different dependencies on 
%black hole mass and accretion rate, and that therefore, as
%Goldschmidt et al.\ (1999) point out, the $R^*$ parameter
%is of limited use for characterising the radio properties of quasars.

\section{Interpretation}

Laor (2000) has suggested that quasars can be characterised by three 
parameters, namely black hole mass, the ratio of bolometric luminosity 
to the Eddington luminosity $L/L_{\rm Edd}$, and orientation. By 
excluding the flat-spectrum quasars from our analysis and 
utilising the $R_{\rm c}$ -- FWHM H$\beta$ correlation we have 
eliminated orientation as a significant variable, leaving just $M_{\rm bh}$ 
and $L/L_{\rm Edd}$. 

\begin{figure}[ht]
\plotone{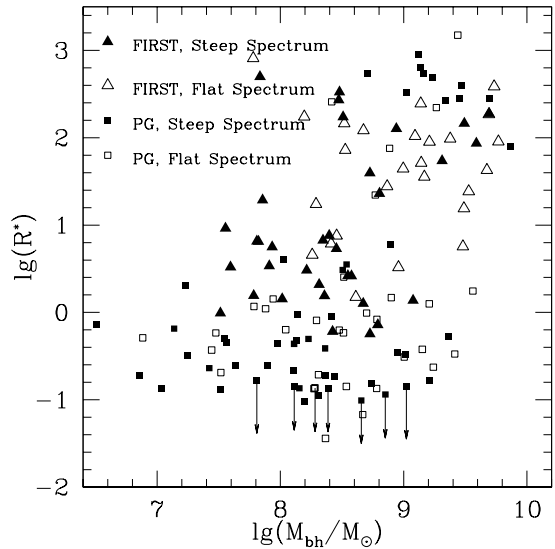}
\caption{Radio-loudness parameter, $R^*$, against black hole mass}
\end{figure}

\begin{figure}[ht]
\plotone{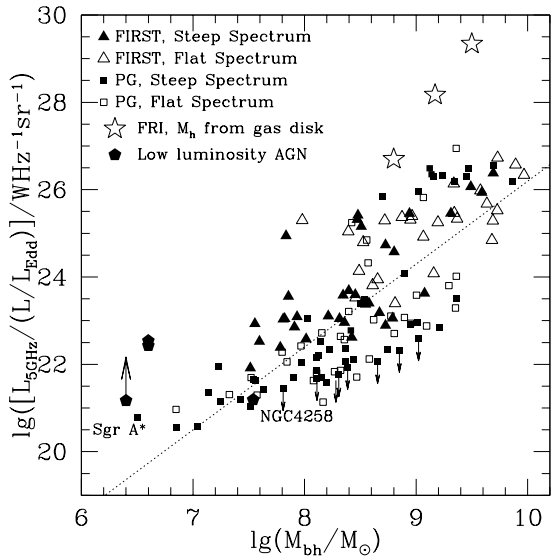}
\caption{Radio luminosity divided by $L/L_{\rm Edd}$ (to account approximately 
for the accretion-rate dependence of radio-luminosity), against black hole
mass. The dotted line is the best fit to the PG and FBQS quasars.}
\end{figure}

To disentangle these dependences, we fitted an equation of the form:
\[ {\rm lg}L_{\rm 5GHz} = a\, {\rm lg}w + b\, {\rm lg}L_B + c \]
to the data for the steep-spectrum PG and FBQS quasars using least-squares, 
where $w$ is the velocity FWHM of H$\beta$. (For objects with only limits on 
their radio luminosity we set the radio luminosity equal to the limit.)
By writing the 
derived quantities $M_{\rm bh}$ and $L/L_{\rm Edd}$ in terms of the
observables $w$ and $L_{\rm B}$ and the best-fit values of $a$ and 
$b$ we obtained the following empirical relation for the radio luminosity:
\[{\rm lg}L_{\rm 5GHz} = 1.9\;{\rm lg}M_{\rm bh} + 1.0\;{\rm lg}(L/L_{\rm Edd})
+ 7.9\, .\]
Corroborating evidence for an accretion rate dependence comes from studies of 
the correlation of radio and optical (continuum and/or emission line) 
luminosity for radio-selected AGN. 
As Fig.\ 2 shows, the range in black hole masses for radio-loud objects is 
very small. Thus the correlation of quasar optical luminosity with 
radio luminosity for quasars selected on the basis of bright radio 
emission (e.g.\ Serjeant et al.\ 1998), must be predominately a correlation
of $L/L_{\rm Edd}$ with radio luminosity (Willott et al.\ 1999). Zirbel 
\& Baum (1995) measure the logarithmic slope of this 
correlation to be 0.28 for FRI sources and 0.75
for FRII sources, and Jarvis et al.\ (2000) measure an even steeper slope for 
luminous high redshift FRII radio sources, close to our estimate of 1.0.

In Fig.\ 4 we plot $L_{\rm 5GHz}/(L/L_{\rm Edd})$ versus $M_{\rm bh}$. The 
correlation is indeed marginally 
tighter than that in Fig.\ 2; the Spearman rank correlation coefficient for 
the steep spectrum objects in Fig.\ 4 is 0.74, compared to 0.59 in Fig.\ 2.
The slope of the correlation, as measured using {\sc emmethod}, is 
$1.9 \pm 0.2$ (although the slope appears to steepen towards 
higher $M_{\rm bh}$, we have not attempted a more complicated fit due to the 
large scatter and uncertainty over starburst contributions at low 
$L_{\rm 5 GHz}$). The fact that the FRI sources and SgrA* plot above the 
correlation in Fig.\ 4 suggests that the accretion rate dependence 
of radio luminosity may weaken at the $L/L_{\rm Edd}\sim 10^{-5}$ 
accretion rates typical of FRI radio sources, consistent with the 
smaller logarithmic slope of the radio-optical correlation seen for 
FRI sources. 
Instead of reflecting a change in accretion-rate dependence, however, 
this weakening may alternatively be explained by a decrease in radiative 
efficiency of these systems at low accretion rates, as predicted by 
models of advection-dominated accretion flows (e.g. Narayan 1996), which 
would lead us to underestimate the true accretion rate. 
The $M_{\rm bh}$ dependence is similar to that found by 
Franceschini et al.\ (1998) for the much less luminous radio sources in nearby
galaxy nuclei
($L_{\rm 5GHz} \propto M_{\rm bh}^{2.5}$). The difference in slope and 
normalisation to the quasar $ L_{\rm 5GHz}$ -- $M_{\rm bh}$ relation 
noted by Laor (2000) is probably explicable in terms of lower accretion 
rates, and a different dependence of $L/L_{\rm Edd}$ on $M_{\rm bh}$ in 
these objects.

\section{Comparison with models}

Energy in the form of magnetohydrodynamic winds and/or a Poynting flux 
can, in principle, be extracted from either spinning black 
holes (Blandford \& Znajek 1977) or accretion disks
(e.g.\ Blandford \& Payne 1982). 
Both the black hole and disk model predictions for the 
power output, $L_{\rm w}$, take a similar form, scaling with 
the square of the black hole radius (i.e.\ with $M_{\rm bh}^2$, 
for a Schwartzschild black 
hole), multiplied by the poloidal magnetic field in the inner disk/black hole 
region, $B_p$, squared (e.g.\ Livio, Ogilvie \& Pringle 1999).

A full model for $B_p$ does not currently exist, though
equipartition assumptions suggest 
$B_p^2 \stackrel{\propto}{_{\sim}} 1/M_{\rm bh}$ 
(e.g.\ Ghosh \& Abramowitz 1997), implying $L_{\rm w} \propto M_{\rm bh}$. 
The accretion rate dependence probably arises through its influence on 
$B_p$. The models of Ghosh \& Abramowitz (1997) do not predict 
a strong accretion-rate dependence at $L/L_{\rm Edd}\sim 0.1$, typical for our 
quasars, but those of Meier (2001) do.

Another possible source of non-linearity in the $L_{\rm 5GHz} - M_{\rm bh}$ 
relation arises from
the conversion of radio jet power, $Q$ (assumed $\propto L_{\rm w}$) 
to radio luminosity in the extended lobes of the radio source. If the minimum 
energy assumption is valid, $L_{\rm 5GHz}\propto 
Q^x$, where $1.2<x<1.75$, depending on the radio source model
(Miller et al.\ 1993). The case of $x=1.75$ corresponds to a 
radio source in pressure balance with its external medium, whereas a purely 
ram-pressure confined, supersonically-expanding FRII radio source has 
$x\approx 1.2$. Models appropriate to the low-luminosity radio sources
of radio-quiet and radio-intermediate quasars have yet to be 
published, but even so it seems hard to reproduce the observed 
$M_{\rm bh}$ dependence solely by this mechanism.

Are other conditions required to produce powerful radio jets? The standard 
deviation in $L_{\rm 5GHz}/(L/L_{\rm Edd})$
about the best-fit line in Fig.\ 4 is 1.1 dex, about what one would expect 
given the $\approx 0.5$ dex accuracy of black hole 
mass estimates from H$\beta$ widths (Laor 1998), and the value of the slope. 
Therefore from 
our study alone there is no compelling statistical evidence for an additional
factor being necessary to produce radio jets. However, the large scatter
may well be hiding contributions from physically important effects,
such as black hole spin and radio source environment. 
A related question is whether there are any actively accreting
high mass black holes with very low radio luminosities in Fig.\ 4. This is 
best addressed using the PG quasars, whose selection criteria are 
radio-independent. There does indeed appear to be a genuine deficiency
(Laor 2000), suggesting that the other variables affect radio 
luminosity by a factor of $\stackrel{<}{_{\sim}} 1000$.

\section{Summary}

A plot of black hole mass versus radio luminosity for quasars 
from the FBQS and PG sample shows a continuous variation of radio luminosity 
with black hole mass for radio luminosities ranging from the traditional 
radio-quiet to radio-loud. Fitting the data for both a black hole mass and 
accretion-rate dependence, we find that radio luminosity scales 
approximately as:
\[L_{\rm 5GHz} \propto M_{\rm bh}^{1.9\pm 0.2} (L/L_{\rm Edd})^{1.0}\]
at accretion rates of $L/L_{\rm Edd}\sim 0.1$. At lower accretion 
rates, we suggest an apparent weakening of the accretion rate 
dependence, to $L_{\rm 5GHz} \propto (L/L_{\rm Edd})^{0.3}$
at $L/L_{\rm Edd}\sim 10^{-5}$ (or a corresponding decrease in the 
radiative efficiency). The radio and optical 
luminosity of quasars have quite different dependences on $M_{\rm bh}$, which 
may explain why the radio luminosity distribution of quasar surveys
differ so dramatically depending on whether or not selection is based
on detection in the radio.

%The strong black hole mass dependence of radio luminosity means that 
%all powerful radio sources have massive black holes. We can thus use 
%measurements of the radio source luminosity function to 
%directly constrain the number density of actively-accreting massive black holes
%at high redshift, independent of dust obscuration. Studies of 
%radio source host galaxies, which have the luminosities
%of giant ellipticals out to $z>2$ (e.g.\ Lacy, Bunker \& Ridgway 2000), 
%suggest that massive black holes continue to be hosted by massive 
%galaxies out to high redshift, and that radio selection can therefore 
%be used to find some of the most massive galaxies at early epochs.

\acknowledgments

We thank Mike Brotherton and Julian Krolik for helpful discussions, 
Ari Laor for a helpful referee's report, and the many  
contributors of spectra to the FBQS database. This work was performed under
the auspices of the U.S.\ Department of Energy by the University of 
California Lawrence Livermore National Laboratory under contract No.\ 
W-7405-Eng-48, with support from NSF grants AST-98-02791 and AST-98-02732.

%% Generally speaking, only the figure captions, and not the figures
%% themselves, are included in electronic manuscript submissions.
%% Use \figcaption to format your figure captions. They should begin on a
%% new page.

%\clearpage

\end{document}